\font\tenit=cmti10
\font\elevenit=cmti10 scaled\magstep 1
\font\elevenbf=cmbx10 scaled\magstep 1
\font\elevenrm=cmr10 scaled\magstep 1

\documentstyle[12pt]{article}
\title{  One-dimensional nonequilibrium
kinetic Ising models with branching annihilating random walk}
\author{\sl Nora Menyh\'{a}rd\\ {\tenit  Research Institute for Solid State
Physics,
 H-1525 Budapest,P.O.Box 49, Hungary}\\
E-mail:menyhard@power.kfki.szfki.hu\\}
\date{  }
\begin{document}
\maketitle
\begin{abstract}
Nonequilibrium kinetic Ising models evolving under the competing effect
of spin flips at zero temperature and nearest neighbour spin exchanges
at $T=\infty$ are investigated
numerically from the point of view of a phase transition.
Branching annihilating
random walk of the ferromagnetic domain boundaries determines
the steady state of the system for a  range
 of parameters of the model. Critical exponents obtained
by simulation are found to agree, within error, with those in
Grassberger's cellular
automata.
\end{abstract}
\nonumber
\newpage
\section{ Introduction }
\medskip
\elevenrm
\baselineskip=14pt

Nonequilibrium kinetic Ising models , in which the steady
state is produced by kinetic processes in connection with heat
baths at different temperatures have been widely investigated[ 1 ]
Most of these studies, however, are concerned with the effects
the nonequilibrium nature of dynamics might exert on phase transitions
driven by  temperature.

A different line of investigating nonequilibrium
phase transitions has been  via
branching annihilating random walk (BAW)  processes. Here
 particles chosen at random carry out random walk (with probability $p$)
with annihilation
upon meeting.  The increase of particles
 is ensured through production of $n$ offsprings with
probability $1-p$. It has been shown
by Sudbury[2] , that the $n=2$ BAW in one dimension leads to extinction
for any $p>0$. Otherwise a phase transition occurs for finite p,
which is in the same universality class as directed percolation (DP)
if $n=odd$, while the critical behavior is different for $n=even$[3 ].

Grassberger et al.[ 4 ] studied probabilistic cellular automata models in
one dimension involving the processes $k\rightarrow3k$ and
$2k\rightarrow0$ ($k$ stands for kink), very similar to BAW
with $n=2$. These models, however,
do show a phase transition and both time-dependent and steady-state
simulations have resulted in non-DP values for the relevant
critical exponents [5 ].  Quite recently, Jensen [6 ] has reported computer
simulation data according to which  the $n=4$ BAW  is in the same
dynamic universality class as Grassberger's cellular automata.

It is the aim of the present paper
to introduce a class of general nonequilibrium kinetic Ising models
(NEKIM) with combined spin flip dynamics at $T=0$ and spin
exchange dynamics at $T=\infty$ in which,
for a range of parameters of the model,  Grassberger-type transition
takes place. The advantage over the cellular automaton formulation is that
in NEKIM
the rates of random walk, annihilation and kink-production processes can
separately be controlled. This circumstance leads to the understanding
of the $n=2$ BAW result which occurs in our model as a special
case.
Results of computer simulations are presented for different
critical exponents and scaling function from random as well as
1-kink initial states.
\vglue 0.5cm
\section{ The model }
\medskip
\elevenrm
\baselineskip=14pt
The model we will investigate here is a one-dimensional kinetic Ising
model evolving by a combined spin-flip and spin-exchange dynamics
with the restriction that spin flip transition probabilities satisfy
detailed balance for  the equilibrium state of the (ferromagnetic) Ising
model at temperature $T=0$ while
the spin exchanges are random ($T=\infty$).

A general form of the Glauber spin-flip transition rate in one-dimension
for spin $s_i$
sitting at site $i$ is [7] ($s_i=\pm1$):
\begin{equation}
W_i = {1\over{2\Gamma}}(1+\delta s_{i-1}s_{i+1})\left(1 -
{\gamma\over2}s_i(s_{i-1} + s_{i+1})\right)
\end{equation}
where $\gamma=\tanh{{2J}/{kT}}$ , with $J$ denoting the coupling constant in
the Ising Hamiltonian
 $\Gamma$ and $\delta$ are further parameters which can
 also depend on temperature. There are
 three independent rates:
\begin{eqnarray}
w_{same}&=&{\frac1{2\Gamma}}(1+\delta)(1-\gamma)\\
w_{oppo}&=&{\frac1{2\Gamma}}(1+\delta)(1+\gamma)\\
w_{indif}&=&{\frac1{2\Gamma}}(1-\delta),
\end {eqnarray}
where the suffices $same$ etc. indicate the three possible
neighborhoods of a given spin($\uparrow\uparrow\uparrow, \downarrow\uparrow
\downarrow and \uparrow\uparrow\downarrow$,respectively).
In the following $T=0$ will be taken,
thus $\gamma =1$,  $w_{same}=0$
and $\Gamma$, $\delta$ are constants to be varied.
 In this limit the three best-known
single spin flip kinetic Ising models correspond to the following
choices of parameters :\\
a). Glauber model [8 ]: $\Gamma=1$, $\delta=0$\\
b). Metropolis model[9 ]: $\Gamma=2/3$, $\delta=-1/3$\\
c). Haake-Thol model[10 ]: $\Gamma=2$, $\delta=1$ .

The spin-exchange transition rate of neighbouring spins, originally
introduced by Kawasaki[7 ] to situations with conserved magnetization,
can be given, for $\gamma =0$, as:
\begin{equation}
w_{ii+1}={1\over2}p_{ex}[1-s_is_{i+1}]
\end{equation}
where $p_{ex}$ is the probability of spin exchange.

The transition probabilities (2)-(4) are responsible for basic elementary
processes of kinks in the  usually expected ferromagnetical ordering
of Ising spins. A kink separating two domains can carry out random walks with
probability
\begin{equation}
p_{rw}\propto 2w_{indif}={1\over\Gamma}(1-\delta)
\end{equation}
while two kinks getting into neighbouring
positions will annihilate with probability
\begin{equation}
p_{an}\propto w_{oppo}={1\over\Gamma}(1+\delta)
\end{equation}
($w_{same}$ is responsible for creation of kink
pairs inside of ordered domains at $T\not=0$).

In case of the spin exchanges, which also act only at phase boundaries,
the process of main importance here is that a kink can produce two
offsprings by the next time step with probability
\begin{equation}
p_{k\rightarrow3k}\propto{p_{ex}}
\end{equation}
The abovementioned three processes compete, and it depends on the
parameters $\Gamma$, $\delta$ and $p_{ex}$ what the result of this
competition will be.
It is important to realize that  the process $k\rightarrow3k$
can  develop into propagating offspring production only if
$p_{rw}>p_{an}$, i.e. the new kinks are able to travel on the average
some
lattice points away from their place of birth an can thus avoid
immediate annihilation. It is seen from  the above definitions
that $\delta<0$ is  necessary for this to happen.
In the opposite case the only effect of the $k\rightarrow 3k$
process
on the usual Ising kinetics is to soften domain walls.
As a matter of fact, in  earlier investigations
(though from a different point of view) of the competing
Ising kinetics in one dimension the choices of parameters
correspond to the latter situation [11].

\vglue .5cm
\section{Phase boundary, critical exponents}
\medskip
\elevenrm
\baselineskip=14pt

We have considered a simplified version of the above model
by keeping only two parameters instead of three by imposing
the condition
\begin{equation}
 p_{k\rightarrow3k}=1-(p_{rw}+p_{an}).
\end{equation}
In the plane of parameters $p_{ex}$ and $ 1-\delta={2p_{rw}/{p_{an}}\over
{1+p_{rw}/p_{an}}}$ the phase diagram shown on Fig.1. has been obtained
by computer simulation. The initial state has been random with zero
average magnetization . Application of our nonequilibrium rule
then corresponds to quenching from $T=\infty$ to $T=0$ with
subsequent exchanges ($T=\infty$) at each time step.
 The line of phase transitions separates
two kinds of steady states reached by the system for large times:
in the Ising phase the system orders while the active phase is
disordered from the point of view of  the underlying spins.
The cause
of disorder is the steadily growing number of kinks with time.
Kink-multiplication is here a branching
process, as explained above, in a sense similar to directed percolation
while in the usual case of Ising  phase transition  with Glauber kinetics
 kinks are created pairwise
inside of ordered domains at nonzero temperature.

 Exponent $\alpha$, used for determining the phase boundary, as well as
 other exponents occuring below are defined
in the framework of the scaling considerations presented by
Grassberger [5].
 The branching increase of kinks becomes obvious when starting
from a single kink state at $t=0$.
The scaling form for the density $\rho{(x,t,\epsilon)}$ of the kinks
has been supposed in the form[5]
\begin{equation}
\rho{(x,t,\epsilon)}\propto{t^{-\alpha}}{\phi(\epsilon{x^{1/{\nu}_\bot}},
\epsilon{t^{1/{\nu}_\Vert}})}
\end{equation}
Here $\epsilon$ measures the deviation from the critical probability
at which the branching transition occurs, $\nu_{\bot}$ and
$\nu_{\Vert}$ are exponents of coherence lengths in space and time
directions, respectively. $\phi(a,b)$ is analytic near $a=0$ and
$b=0$.
Using (10) the following relations
can be deduced.

The average particle number $N(t,\epsilon)$ grows in the active phase as
\begin{equation}
 N{(t,\epsilon)} \propto{{t}^\eta{\psi(\epsilon t^{1/{\nu}_\Vert)}}}.
\end{equation}

 The RMS size of the cluster growing from
the single-kink initial configuration is given by
\begin{equation}
{<{(x-<x(t)>)}^{2}>}^{1/2}\propto{<x_{max}(t)-x_{min}(t)>}\propto{t^{Z/2}}
\end {equation}
where $Z=2\nu_{\bot}/{\nu_{\Vert}}$.

When starting from a random initial state
the exponent $\beta$ characterizes the growth of the average kink
density in the active phase:
\begin{equation}
\rho(\epsilon)=\lim_{t \rightarrow \infty}\rho(x,t,
{\epsilon})\propto{{\epsilon}^\beta}
\end{equation}
while the decrease of density at the critical point is given by
\begin{equation}
n(t)\propto{t^{-\alpha}}.
\end{equation}
The exponents are connected by scaling laws: $\beta=\nu_{\Vert}\alpha$,
$\alpha={Z/2}-\eta$.

The phase boundary shown on Fig1. has been identified using (14).
It is worth noting that the critical point marked on the phase boundary
has been chosen in a region where the width of the critical region
is relatively small. The rest of the critical exponents have been
measured only around this point :${\frac1{\Gamma}_{c}}=.35$,
$\delta_{c}=-.4\pm.01$ and
${(p_{ex})}_{c}={1-\frac2{{\Gamma}_{c}}}=.3\pm.01$.
The deviation  from the critical
point has been chosen in the direction of $\delta$.

The critical region is especially wide near the two ends of the
curve. Moreover, near $p_{ex}=1.0$ where we get close to
$\delta=0$ (Glauber case), it is very hard to determine $\alpha$:
the exponent grows slowly with time from a nearly zero value
at small times. Several  runs with different values of the parameters also
without the restriction (9),
have been performed in this Glauber limit all showing that
the $\delta=0$ case for all $\Gamma$, $p_{ex}$ values, remains
Ising-like: the exponent $\alpha$ tends to the value $\alpha=.5$
for large times.
The accuracy of our simulations does not allow us to make
a quite definite statement,
nevertheless it seems very likely that the asymptote of the phase boundary
is $\delta=0$ for $p_{ex}=1.0$. Therefore the conjecture that
$p_{rw}>p_{an}$  is a necessary condition for a BAW-type phase transition to
occur seems to be supported by  the present simulations.

The value of $\alpha$ provided by our simulations  agrees, within
error, with that obtained
by Grassberger[4,5] in a one -dimensional elementary cellular automaton
model with special added noise $p$. The same is true also for the rest
of the critical exponents reported below.
Fig.2 shows the density of kinks
  in the
active phase for different values of the deviation from the critical point
$\epsilon=\mid{\delta-\delta_c}\mid$, starting from  random initial
states. At $\epsilon=0$ the power law behaviour
(14), is seen with $\alpha=.27\pm.02$. $\beta$ has been obtained
directly,  according to eq.(13), by taking the level-off values
of $n(t) $, some of which can be read off from Fig.2.
 Though this is known to be a fairly inaccurate way to
determine $\beta$ [5], we have found -though only over one decade in
$\epsilon$ - the reasonable value $\beta=.8\pm.08$.

On Figs.3 through 5 results of simulations starting from one-kink initial
states are seen. Data for $N(t,\epsilon)$ at $\epsilon=0$ are presented
on Fig.3; the straight line on the log-log plot leads to $\eta=.30\pm
.02$. Fig.4 shows the growth of the average cluster size, eq.(10),
again on log-log scale, yielding ${Z/2}=.57\pm.01$. Supposing the scaling
law  $\beta=\nu_{\Vert}\alpha$ to be valid we get $\nu_{\Vert}=2.9$.

Using the values obtained for the exponents, we have checked the
consistency of the data within the scaling framework by measuring
the scaling
form, eq.(11). The result is seen on Fig.5 for some values of
$\epsilon$, nevertheless also only inside of a decade, because
of computational difficulties. The second scaling law :$\alpha={Z/2}-
\eta$ is also fulfilled with the exponents found in our simulations.
\vglue 0.5cm
\section{ Connection with BAW }
\medskip
\elevenrm
\baselineskip=14pt

The definition of BAW is the following. The process is taking place in two
steps: a). a particle chosen at random moves to a nearest neighbour
 place also randomly
chosen, with probability $p$ and b). a particle chosen at random produces
n offsprings on neighbouring sites with probability $1-p$. If any of
the neighbouring places in question is already occupied then annihilation
takes place. The $k+k\rightarrow0$ process is prescribed to occur
with probability unity. As mentioned in the introduction, no phase
transition takes place in the $n=2$ case.

We can bring into relation the $n=2$ BAW in 1d and
NEKIM by noticing that in the BAW rule the transition probability
for annihilation is unity. From eq.(7) $p_{an}=1$ leads to
$\Gamma=(1+\delta)$ giving, with eq.(6), $p_{rw}={{1-\delta}\over{1+\delta}}$.
{}From the trivial correspondence
\begin{eqnarray}
p_{rw}&\cong& p\\
p_{k\rightarrow3k}&\cong& 1-p
\end{eqnarray}
while changing $p$,  the parameters of the corresponding NEKIM models
take the
 values from $\delta=1$, $p_{ex}=1$, $\Gamma=2$ at $p=0$ through
values $\delta>0$, $p_{ex}<1$, $\Gamma<2$ to $\delta=0$, $p_{ex}=0$
, $\Gamma=1$ at $p=1$. In these regions, however, no phase transition
has been found to take place, all steady states are Ising-like.
\vglue 0.5cm
\section{ Discussion }
\medskip
\elevenrm
\baselineskip=14pt

 Non-thermal phase transition has been found in a family of one-
dimensional kinetic Ising models evolving by a combined spin-flip and
spin-exchange dynamics. The single spin flip transition rate is
supposed  in its most general form.
It contains three parameters of which the temperature is
fixed at zero where the remaining two control the probability of
random walk and annihilation. The nonzero transition rate of spin
exchange leads to $k\rightarrow 3k$ -type kink production and
provided RW overwhelms annihilation, increasing rate of kink
production results in a phase transition from Ising-like to active
phase. Numerical simulations have given evidence that the model is
in the same universality class as Grassberger's  cellular
automata models[4,5], where a certain stochastic element
(with probability p) of a simple 1d CA rule
drives the transition.
For $p=0$ the rule is deterministic and
the steady state is infinitely degenerate, when starting from a random
initial situation the system settles in one of these states after a few
iteration steps. The corresponding kinetic Ising model is the
Haake-Thol model in the $T\rightarrow 0$ limit
mentioned in section 2. Increasing $p$ up to $p_c$ in the CA model,
could correspond to a line on our phase diagram (Fig 1)
starting at the origin and ending at some intermediate point of
the transition line. It would be difficult, however,
 to make a closer (quantitative)
correspondence between the two kinds of models since in CA
 this single parameter p is responsible for random walk as well as
$k\rightarrow3k$ in a complicated,nonlinear way.
   To find correspondence between our model and the two offspring BAW
is more
straightforward. For this aim restriction (9) had to be lifted since
$p_{an}=1$ is taken in BAW rules and then eq (9) would lead to negative
probabilities. According to the correspondence between BAW and NEKIM
it is clear that
a necessary condition to find a phase transition is $p_{an}<1$.

\vglue .5cm
\section{ References }
\medskip
\elevenrm
\baselineskip=14pt
\begin{enumerate}
\item{\elevenit  R\'acz Z and Zia K P }Phys.Rev.E {\elevenbf 49} 139
(1994) and references therein\\
\item{\elevenit Sudbury A.} Ann. Probab. {\elevenbf 18} 581 (1990)\hfill
\item{\elevenit Jensen I} J.Phys. A:Math. Gen. {\elevenbf 26} 3921 (1993)\hfill
\item{\elevenit Grassberger P, Krause F and von der Twer T}\\
J.Phys.A:Math.Gen.{\elevenbf 17}(1984)L105\hfil
\item {\elevenit Grassberger P} J.Phys. A:Math.Gen.
{\elevenbf 22} (1989) L1103\hfil
\item{\elevenit Jensen I} University of Melbourne Preprint, May 1994\\
\item  see e.g. {\elevenit Kawasaki K:}
 Phase Transitions and Critical Phenomena,Vol.2.,\hfil
 ed.{\elevenit Domb C and Green M S}
 (New York: Academic, 1972) p.443\hfil
\item {\elevenit  Glauber R J } J.Math.Phys.{\elevenbf 4}
 (1963) 191\hfill
\item {\elevenit Metropolis N , Rosenbluth A W, Rosenbluth M N,
 Teller A H and Teller E,}\\ J.Chem.Phys. {\elevenbf 21} (1953) 1087\hfil
\item {\elevenit  Haake F  and Thol K,} Z.Physik B:
 Condensed Matter,{\elevenbf 40} (1980) 219
\item {\elevenit Droz M, R\'acz Z and Schmidt J}\\
Phys.Rev. {\elevenbf A39} (1989) 2141\hfil
\end{enumerate}
\newpage
\medskip
{\elevenbf Figure captions}\\
\medskip
\elevenrm
\baselineskip=14pt

{\elevenbf Fig 1}  Phase diagram of the two-parameter model. The phase
boundary has been obtained by measuring $n(t)$, the density of
kinks, starting from a random initial distribution and locating the
phase transition points by $n(t)\propto t^{-\alpha}$ with
$\alpha=.27\pm.04$. Typically the  number of lattice points has been
$l=2000$  and averaging over 500  independent runs has been performed.
Checks for unimportance of finite size effects have been carried out.
The dotted line with arrow indicates the critical point and the
direction of deviation from it along which further critical
characteristics have been determined.
\\
\\
{\elevenbf Fig 2} Density of kinks as a function of time on a
log-log plot for different values of $\epsilon=\mid \delta-\delta_{c}\mid$,
where $\delta_{c}=-.4\pm.01$ is the critical value of parameter $\delta$
at the chosen critical point. $p_{ex}$ is kept fixed at its critical value.
Lattice sizes : $2000-8000$, number of independent runs :$ 400-800$.
\\
\\
{\elevenbf Fig 3} The growth of the average particle number at
the critical point when starting from a single-kink initial state.
The number of independent runs in the averaging: 900. The
scale is double logarithmic.
\\
\\
{\elevenbf Fig 4} Average distance between rightmost and leftmost
kinks at the critical point on a double logarithmic scale.
Average over 900  independent runs has been carried out.
\\
\\
{\elevenbf Fig 5} Scaling function $\psi(\epsilon{t}^{1/\nu_{\Vert}})$
of eq.(11) for different values of $\epsilon$. Initial state: single-kink,
number of independent runs in averaging : 500.
\end{document}